\documentclass{article}

\usepackage{amsmath}

    \usepackage[preprint]{neurips_2024}

\usepackage[utf8]{inputenc} 
\usepackage[T1]{fontenc}    
\usepackage{hyperref}       
\usepackage{url}            
\usepackage{booktabs}       
\usepackage{amsfonts}       
\usepackage{nicefrac}       
\usepackage{microtype}      
\usepackage{xcolor}         

\usepackage[inline]{enumitem}
\usepackage{tcolorbox}
\usepackage{listings}
\usepackage{subcaption}
\usepackage{cleveref}
\usepackage{xspace}
\usepackage{amssymb}
\usepackage{inconsolata}
\newcommand{\compiler}{\textsc{LLMLift}\xspace}

\crefname{section}{Sec.}{Sec.}
\crefname{figure}{Fig.}{Fig.}
\Crefname{figure}{Fig.}{Fig.}
\crefname{table}{Tab.}{Tab.}
\Crefname{table}{Tab.}{Tab.}
\Crefname{equation}{Eq.}{Eq.}
\crefname{equation}{Eq.}{Eq.}
\crefname{appendix}{Appendix}{Appendix}

\definecolor{anti-flashwhite}{rgb}{0.95, 0.95, 0.96}
\newcommand{\src}{{\tt S}\xspace}
\newcommand{\trg}{{\tt T}\xspace}
\newcommand{\slang}{$S_{lang}$\xspace}
\newcommand{\tlang}{$T_{lang}$\xspace}
\lstdefinelanguage{java}
{
  morekeywords={for, int, static, class, public, void, if, new, return, RDD, string},
  morecomment=[l]{//},
  morecomment=[s]{/*}{*/},
  morestring=[b]",
  basicstyle=\footnotesize\ttfamily,
  numbers=left, firstnumber=1, numberstyle=\tiny\color{gray},
  showstringspaces=false,
  escapeinside={(*@}{@*)},
  keywordstyle=\color{blue},
  columns=fullflexible,
  showlines=true,
  xleftmargin=0.5cm,
  moredelim=**[is][\color{OliveGreen}]{<_}{_>}
}
\lstdefinelanguage{python}{
    morekeywords={def, return, if, else, len, map, reduce, ite},
    escapeinside={(&}{&)},
    numbers=left, firstnumber=1, numberstyle=\tiny\color{gray},
    xleftmargin=0.5cm,
    keywordstyle=\color{blue},
    basicstyle=\footnotesize\ttfamily,
    showlines=true,
    columns=fullflexible    
}

\lstdefinelanguage{cpp}{
    morekeywords={void, for, int},
    escapeinside={(&}{&)},
    numbers=left, firstnumber=1, numberstyle=\tiny\color{gray},
    xleftmargin=0.5cm,
    keywordstyle=\color{blue},
    basicstyle=\footnotesize\ttfamily,
    showlines=true,
    columns=fullflexible    
}

\setlist[itemize]{leftmargin=*}
\setlist[enumerate]{leftmargin=*}

\newif\ifcomments
\commentsfalse
\ifcomments
    \providecommand{\sahil}[1]{{\protect\color{red}{\bf [sahil: #1]}}}
    \providecommand{\alvin}[1]{{\protect\color{purple}{\bf [alvin: #1]}}}
    \providecommand{\jie}[1]{{\protect\color{teal}{\bf [jie: #1]}}}

\else
    \providecommand{\sahil}[1]{}
    \providecommand{\alvin}[1]{}
    \providecommand{\jie}[1]{}

\fi

\title{Verified Code Transpilation with LLMs}

\author{
Sahil Bhatia \\
UC Berkeley \\
\texttt{sahilbhatia@berkeley.edu} \\
\And
Jie Qiu \\
Duolingo \\
\texttt{jieq@berkeley.edu} \\
\And
Niranjan Hasabnis\\
Intel Labs\\
\texttt{niranjan.hasabnis@intel.com}
\And
Sanjit A. Seshia\\
UC Berkeley \\
\texttt{sseshia@berkeley.edu}
\And
Alvin Cheung \\
UC Berkeley \\
\texttt{akcheung@berkeley.edu}
}

\begin{document}

\maketitle

\begin{abstract}
  Domain-specific languages (DSLs) are integral to various software workflows. Such languages offer domain-specific optimizations and abstractions that improve code readability and maintainability.  
  However, leveraging these languages requires developers to rewrite existing code using the specific DSL's API.
  While large language models (LLMs) have shown some success in automatic code transpilation, none of them provide any functional correctness guarantees on the transpiled code.
  Another approach for automating this task is verified lifting, which relies on program synthesis to find programs in the target language that are functionally equivalent to the source language program. While several verified lifting tools have been developed for various application domains, they are specialized for specific source-target languages or require significant expertise in domain knowledge to make the search efficient. In this paper, leveraging recent advances in LLMs, we propose an LLM-based approach (\compiler) to building verified lifting tools. We use the LLM's capabilities to reason about programs to translate a given program into its corresponding equivalent in the target language.  
  Additionally, we use LLMs to generate proofs for functional equivalence. We develop lifting-based compilers for {\em four different} DSLs targeting different application domains. Our approach not only outperforms previous symbolic-based tools in both the number of benchmarks transpiled and transpilation time, but also requires significantly less effort to build.
\end{abstract}

\section{Introduction}
\label{sec:intro}

Domain-specific languages (DSLs) have gained popularity 
due to their ability to provide optimizations and abstractions that enhance code readability and improve performance in specific domains. Examples of recent DSLs include Spark (distributed computing), NumPy (array processing), TACO (tensor processing), and P4 (network packet processing). With new DSLs emerging for diverse application domains and programming languages, developers often face the task of manually rewriting existing code to incorporate these languages into their existing workflows. This manual rewriting process can be tedious, may introduce bugs into the code, and may fail to preserve the semantics of the starting code. This problem of transforming and compiling code from one programming language to another is called {\em transpilation}. 
The question we address in this paper is: can large language models (LLMs) correctly and automatically perform code transpilation?

A particularly useful form of code transpilation, termed {\em lifting}, involves translating code in a somewhat lower-level, general-purpose language to equivalent code in a DSL. 
Lifting allows developers to port code to DSLs from which efficient code can be generated for special-purpose hardware, such as GPUs, machine learning accelerators, or network processors.
%
Therefore, significant effort has been dedicated to developing tools aimed at automating the task of lifting. 
Rule-based approaches rely on traditional pattern-matching techniques~\cite{mold}; however, describing these rules can be a complex, human-intensive task. An alternative are search-based techniques that leverage advances in {\em program synthesis} (e.g., see~\cite{solar-asplos06,jha-icse10,progsynthesis-bookch17}) and formal verification over the last two decades. The use of verified program synthesis for lifting, termed {\em verified lifting}, involves searching for a program in the DSL and subsequently formally verifying its semantic equivalence to the source program.  Verified lifting has been successfully applied in building compilers~\citep{casper,c2taco,qbs,dexter} for DSLs like Spark, SQL, Halide, and TACO. Contemporary program synthesis approaches can be broadly classified into two categories: {\em symbolic} and {\em neural}. Traditionally, symbolic techniques such as enumerative, deductive, and constraint-based synthesis strategies have been used for implementing the search. More recently, neural networks~\cite{ngst} have been trained and leveraged to accelerate the search process. Despite their successes, both symbolic and neural approaches have common drawbacks:
\begin{enumerate*}[label={\arabic*)}]
\item The synthesizer is customized for each DSL, making them challenging to adapt for new DSLs, and
\item Significant effort is required to design the synthesizer, including domain-specific heuristics for symbolic approaches and the generation of parallel corpora $\langle source, target \rangle$ for ML-based approaches, to enable generalization and scalability for the target DSL.
\end{enumerate*}

Large Language Models (LLMs)~\cite{bert,few-shot} have emerged as a promising approach for tackling complex programming tasks, including code generation, repair,  and testing. However, generating reliable code with formal correctness guarantees with LLMs remains challenging. 
Most work on LLMs either focuses on generating code without correctness guarantees~\citep{starcoder,alphacode,codellama} or separately on producing proof annotations (such as invariants) for given code~\cite{llminv,llminvrank}.  
Additionally, formal verification tools often have their own specialized languages (e.g., SMT-LIB, Dafny) for encoding verification problems and specifications. These languages are typically low-resource in the training datasets of LLMs, making it challenging for the models to generate code in these formal verification languages directly. To leverage LLMs for building VL compilers, we must address two key constraints: generalization to new DSLs and providing correctness guarantees for the generated code.

In this work, we investigate the use of {\em LLMs for verified lifting (VL)}. Our approach, called \compiler, takes inspiration from the core technique of VL, which involves translating the source program to a higher-level intermediate representation (IR) that describes the semantics of the DSL operators. Once the synthesized code is verified, it is then translated to the concrete syntax of the DSL using rewrite rules. 
We leverage the reasoning capabilities of LLMs to translate code from context to an IR. We instruct the model via a prompt to generate code using the operators of the DSL, with Python serving as the IR to encode the semantics of these operators. Python's significant representation in the training datasets of LLMs makes it a suitable choice for this purpose. In addition to generating the DSL program, we also prompt the model to generate a proof of correctness for the program. 
To the best of our knowledge, our approach is the first to leverage LLMs to generate {\em both code and proof annotations} together. To verify the functional equivalence of the generated program to the given source program for all program states, we translate both the generated program and the proof to the syntax of an automated theorem prover. This step ensures that the synthesized code is formally verified and can be trusted to be correct. Our evaluation (section~\cref{sec:approach}) shows that \compiler has significant advantages over traditional search-based symbolic VL-based tools. It solves \textbf{7} more benchmarks, requires substantially less effort in terms of LoC (\textbf{1000$\times$}), and is faster in generating verified code and proofs (\textbf{6$\times$} on average) .

In summary, this paper makes the following novel contributions
\begin{enumerate}[nosep,leftmargin=1.5em,labelwidth=*,align=left]
\item We introduce the first technique for formally-verified code transpilation using LLMs.
\item Our approach uses Python as an IR for code generation, thus eliminating the need for specialized DSL-specific training data or fine-tuning of LLMs.
\item 
Our method eliminates the need for manual encoding of domain-specific heuristics,
thus simplifying the process of verified lifting by reducing the human effort required in traditional techniques.

\item We propose an approach to generate not only the lifted code but also a proof of correctness for the generated code. This integration of LLMs with verification oracles guarantees the correctness of the generated code, a crucial aspect that sets our approach apart from other work on LLM-based code generation.

\item We show the effectiveness of our approach (\cref{sec:experiments}) by constructing compilers for \textbf{four} DSLs spanning various application domains. In terms of accuracy, our LLM-based compilers achieve comparable performance to existing tools and, in some domains, outperforms the prior approaches.

\end{enumerate}
\section{Background}
\label{sec:over}

\begin{figure}
\begin{subfigure}{0.5\textwidth}
\begin{lstlisting}[language=java,basicstyle=\scriptsize\ttfamily]
public class ConditionalSum {	
  public static int sumList(List<Integer> data) {
    int sum = 0;
    for (int i = 0; i < data.size(); i++) {
      int var = data.get(i);
      if (var < 100)
        sum += var;
    }
    return sum;
  }
}
\end{lstlisting}
\caption{Source Code (\src{})} 
\label{fig:src}
\end{subfigure}
\begin{subfigure}{0.35\textwidth}
\begin{lstlisting}[language=python,basicstyle=\scriptsize\ttfamily]
def map(data,f):
  if len(data) == 0: return []
  else:
    return [f(data[0])] + map(data[1:], f)

def reduce(data,f):
  if len(data) == 0: return 0
  else:
    return f(data[0], reduce(data[1:], f))

def ite(a, b, cond):
  if cond: return a
  else: return b
\end{lstlisting}
\caption{Target Language (\tlang{})}
\label{fig:semantic_ops}
\end{subfigure}
\caption{Sequential source code in Java and semantics of DSL in IR}
\vspace{-0.2in}
\end{figure}
We now give an overview and an end-to-end example of verified lifting (VL) 
where we use program synthesis to build a compiler. Given a program (\src{}) in the source language (\slang{}), VL uses a search procedure to find a program (\trg{}) in the target language (\tlang{}) that can be proved to be functionally equivalent to the given source program. 

VL comprises of three phases: 
\begin{enumerate*}[label={\arabic*)}]
    \item Search,
    \item Verification, and
    \item Code generation.
\end{enumerate*}
The key behind VL is to first transpile \src to an user-defined intermediate representation (IR) of the operators in the target language before generating executable code. The IR serves as a {\em functional description} of \tlang and ignores any implementation details. 
Hence, during search phase, \src is \textbf{lifted} to a sequence of operators expressed using the IR. This expression serves as the program summary (PS) which summarizes \src{} using the IR. Subsequently, PS is \textbf{verified} using a theorem prover to check for semantic equivalence with \src{} for all program inputs. If verification succeeds, PS is then translated into the concrete syntax of the target language using simple pattern-matching rules provided by the user to generate executable code. These rules are notably simpler to write compared to a rule-based translator that directly compiles from \slang{} to \tlang{}, as the PS is already expressed using the operators in the target language. 

We demonstrate an example of transpiling a sequential Java program to Spark using VL. Spark provides a high-level API for large-scale distributed data computation. Two key higher-order functions in Spark's API are {\tt map} and {\tt reduce}: {\tt map} applies a given function to each element of a distributed dataset and creates a new dataset, while {\tt reduce} combines the elements of a dataset using a specified associative binary operator, such as summing across the entire dataset.

\Cref{fig:src} shows a sequential source program ({\tt S}). The given \src{} takes a list of integers as input and calculates the sum of all integers in the list that are less than 100. In \Cref{fig:semantic_ops}, we define the semantics of the {\tt map} and {\tt reduce} operators from Spark in Python (the IR). These functions abstract the implementation details of the operators while only capturing the high-level semantics of the operators. Our goal is to find an IR expression sequence of {\tt map} and {\tt reduce} such that it is semantically equivalent to \src{}. Traditional approaches to solving this search problem in VL involve framing it as SyGuS~\cite{sygus} problem. SyGuS is an approach for solving program synthesis problems by specifying constraints and searching for solutions within a defined space. Specifically, a SyGuS problem involves defining a search space that syntactically restricts the space of possible solutions, thereby making the search tractable. Formally, this objective can be stated as $\exists \; T \;  \in \; T_{lang} \mid \forall \; \sigma. \;  S(\sigma) = T(\sigma),$ where T is a program in the target language. For our program in \cref{fig:src}, the synthesis phase would return the following PS (i.e., $T$):  
\begin{lstlisting}[language=python,numbers=none]
reduce(map(data, lambda i : ite(i < 100, i, 0)), lambda a, b: a + b)
\end{lstlisting} 
The expression initially maps each element in {\tt data} to either {\tt i} or 0 based on whether the element {\tt i} is less than 100 or not. Next, it reduces the resulting list by summing up all the elements to return the sum of elements less than 100. Since \src contains a loop, proving equivalence with the generated program requires another predicate called the ``loop invariant.'' A loop invariant is a logical statement that must hold  before and after each iteration of a loop. Intuitively, it captures the essential properties that are preserved while the loop executes. During VL's synthesis phase, we generate both the program summary and any required loop invariant for verification. Verification is done by sending the program summary and loop invariant(s) to a theorem prover. Verifier checks the semantic equivalence between \src{} and the generated program program summaries. 

VL currently uses cvc5 and z3 for this purpose.
 
Once verified, we translate the generated program summary to the concrete syntax of the DSL (Spark) using simple pattern-matching rules, resulting in the following executable code:
\begin{lstlisting}[language=python,numbers=none]
map(lambda i: i if i < 100 else 0).reduce(lambda a, b: a + b)
\end{lstlisting}
We next describe our LLM-based approach can improve the efficiency and scalability of VL's synthesis problem.

\section{LLM-Based Verified Lifting}
\label{sec:approach}

\begin{figure}
\centering

    \centering
    \includegraphics[scale=0.4,trim=2cm 0cm 0cm 0cm]{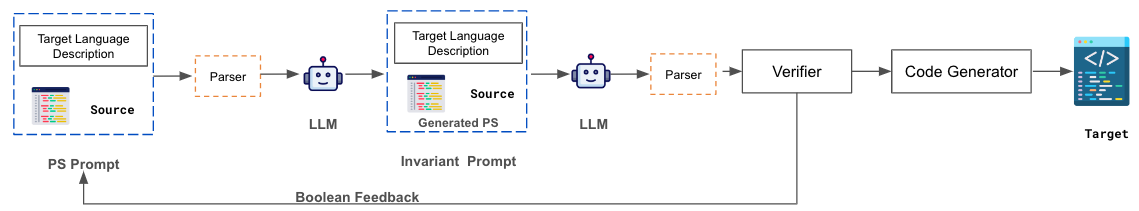}
    \caption{A high-level overview of our \compiler framework for building verified lifting-based tools. }
    \label{fig:llm_framework}
\label{fig:framework}
\end{figure}

We now describe our LLM-based approach for verified lifting. We begin by formalizing VL. Then we give details of how we use LLMs to improve over the classical verified lifting approach.


\subsection{Problem Formulation}


VL's search problem is characterized by three components:
\begin{enumerate}[nosep,leftmargin=1.5em,labelwidth=*,align=left]
    \item \textbf{Specification} ($\phi$): The specification ($\phi$) defines the property that the target program ($\trg{}$) should satisfy. For VL considered in this paper, source and target programs are side-effect free functions of their inputs. Thus, $\phi$ encodes the semantic equivalence of $\trg{}$ to the source program ($\src{}$) for each program input state $\sigma$. The overall correctness condition is:
\begin{equation}
\forall \sigma \; \phi(\sigma, \trg{}, \src{}) \, \doteq \,  
\forall \; \sigma. \; \src{}(\sigma) = \trg{}(\sigma) 
\end{equation}

    \item \textbf{Program Space} ($G$): The program space outlines the set of potential solutions, typically expressed as a context-free grammar $G$. 
    The language of $G$ includes all sequences of operators $ops \in T_{lang}$ applied recursively to terms starting with input variables $\sigma$.
    A target program $\trg{}$ as a program summary PS that is a composition of operators $ops$.
    An example involving the {\tt map} and {\tt reduce} operators is provided in the previous section. 
%
%
In other words, all values returned by \src must be expressed using a combination of operators ($ops$) from \tlang.

    \item \textbf{Search Algorithm} ($A$): This refers to the algorithm used to solve the synthesis problem. Traditional symbolic program synthesis solvers utilize enumerative search, deductive search, and constraint-based approaches~\cite{progsynthesis-bookch17}. 
    Part of the synthesis problem is to generate the invariants $Inv$ that a verifier can use to prove that $\phi$ holds.
    Synthesis tools typically also use a grammar $G_I$ to constrain the space of possible invariants to search over.
\end{enumerate}
In summary, with the target program $\trg{}$ represented as the combination (PS,$Inv$), we define the search problem in VL as:
\begin{equation}
   \exists \; \text{PS} \in G \; \; \exists Inv \in G_I \; \; \forall \sigma \;. \; \phi(\sigma, \text{PS,} Inv, \src{}) \label{eq:vl_spec}
\end{equation}
%
This states that we aim to find a program summary (PS) and invariants ($Inv$) from the defined search space $G,G_I$, using the search algorithm $A$, such that the given specification (functional equivalence with \src{}) holds for all possible program states.

\begin{figure}
\begin{minipage}{0.48\textwidth}
\centering
\includegraphics[scale=0.39]{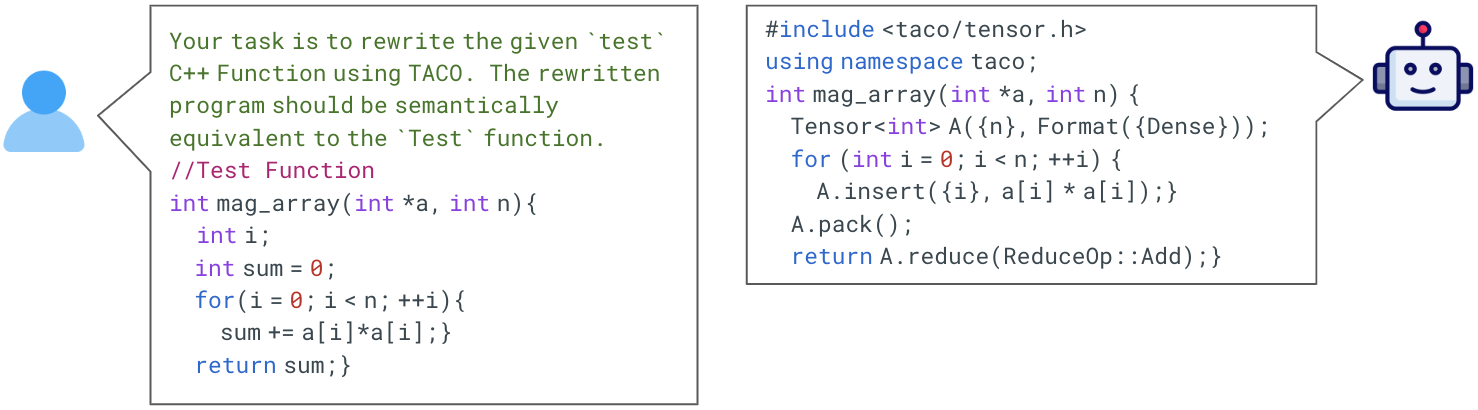}
\caption{End-to-End Lifting Example 
}
\label{fig:e2e}
\end{minipage}\hfill\hfill
\begin{minipage}{0.3\textwidth}
 \centering
    \includegraphics[scale=0.35, trim=0cm 0cm 0cm 0cm]{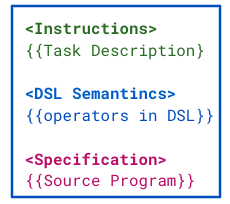}
    \caption{Prompt Structure}
    \label{fig:prompt_struc}
\end{minipage}
\end{figure}

\subsection{LLM-based Verified Lifting}
Traditional approaches to solving the VL search problem rely on symbolic search and manually designed heuristics to make the search effective. Unfortunately doing so is resource intensive and requires domain-specific knowledge.
We explore a new approach by leveraging LLMs.

A naive approach to building a VL-based compiler using LLMs would be to prompt LLMs to translate \slang{} programs directly into \tlang{}. However, this approach has the following shortcomings:
\begin{enumerate}
    \item VL-based compilers require that the \tlang{} candidates generated during the search phase be  functionally equivalent to the input \slang{} program. This is a strong requirement that current LLMs are unable to satisfy.
    \item 
    Contarary to general purpose languages (such as Python), domain-specific languages (DSLs) are not widely used. Unsurprisingly, we find that LLMs struggle to generate code in languages that are insufficiently represented in their training data.
\end{enumerate}

\cref{fig:e2e} shows an example of instructing GPT-3.5 to translate code in an end-to-end manner. We instruct the model to translate a C++ function to TACO (a tensor processing DSL)~\cite{taco}, and the model fails to generated the expected einsum representation. Instead, the model outputs a completely incorrect solution by hallucinating non-existent TACO functions. This problem is even more prominent for new DSLs that the model might have never seen in the training dataset.

To address these challenges, we leverage VL's key idea of {\em  transpiling to an IR rather than directly to the concrete syntax of \tlang{}}. Specifically, we observe that Python is one of the well-represented programming languages in the training dataset of popular LLMs~\cite{starcoder}, and, consequentially, these LLMs understand semantics of Python programs well. We exploit these observations by leveraging Python as the IR to define semantics of DSL operators, an example is shown in~\cref{fig:semantic_ops}.

In~\cref{fig:framework}, we show our LLM-based approach where we apply LLMs to generate the program summaries and invariants in the IR. Each generated summary is then checked for correctness using a program verifier. After verification, we convert the program summaries into the concrete syntax of \tlang{} through simple pattern-matching rules as in traditional VL. Our approach uses LLMs with a few-shot learning framework that we describe next.

\subsection{Few-Shot Learning Approach}

LLMs have demonstrated few-shot reasoning capabilities~\cite{few-shot}. Few-shot reasoning allows LLMs to generalize their understanding to new tasks by leveraging a small set of similar examples. This allows them to extend their reasoning capabilities to tasks without requiring explicit training or fine-tuning for those specific tasks. We propose leveraging the few-shot reasoning capabilities of LLMs for verified lifting as fine-tuning existing LLMs for each new DSL is often infeasible due to the lack of extensive training data and the rapid pace at which new DSLs are developed. The effort required to collect, annotate, and preprocess DSL-specific training data for fine-tuning can be substantial, making it impractical to adapt LLMs to each new DSL.

As described in \cref{sec:over}, VL generates candidates in an IR that abstracts away low-level implementation details of the operators in \tlang{}. The objective, as defined in \cref{eq:vl_spec}, is to find PS and Inv expressed using operators from \tlang{} such that $\phi$ holds. We leverage the few-shot reasoning capability by providing the models with the semantics of operators from the target language (\tlang{}) using an IR. By exposing the LLMs to these semantics, we enable them to use their reasoning capabilities over code to generate both the PS and invariants in the IR.

In \cref{fig:prompt_struc}, we illustrate the high-level prompt structure we use to generate the PS and Invs. The prompt consists of the following components:  

\begin{enumerate}
[nosep,leftmargin=1.5em,labelwidth=*,align=left]
    \item \textbf{Task Instruction.} We instruct the model using a natural language to translate \src{} using only the specified DSL operators.
    \item \textbf{DSL Operators.} We specify the semantics of all operators from \tlang{} using Python and include it in the prompt. Python is chosen as our IR due to \begin{enumerate*}
    \item its widespread use across domains,
    \item its concise and expressive nature, making the representation readable and straightforward and,
    \item its significant representation in code datasets used for training LLMs~\cite{starcoder}.
    \end{enumerate*} 
    \item \textbf{Specification.} While symbolic techniques often rely on approaches like test cases, bounded model checking, and Hoare logic~\cite{hoare-logic} for defining specifications, the natural language interface of LLMs offers flexibility in using various specifications and combining different forms. Given that LLMs are primarily trained on raw source code and may not have encountered other forms of specification during training, we directly use the source program (\src{}) as the specification in our prompt.
\end{enumerate}

We split the generation of PS and $Inv$ into a two-phase process by first asking the LLM to generate the PS and then inferring invariants corresponding to it. For generating PS we use zero-shot setting while for $Inv$(s) we use one-shot prompt.Due to space constraints, we show an instantiation of the prompt structure shown in~\cref{fig:prompt_struc} in~\cref{sec:prompt}. When prompted, the model generates the following PS for our example code shown in~\cref{fig:src}:
\begin{lstlisting}[language=python,numbers=none]
reduce(map(data, lambda i : ite(i < 100, i, 0)), lambda a, b: a + b)
\end{lstlisting}
\vspace{-0.1in}
To ensure that the generated candidates follow the DSL operators defined in the prompt, we use a parser to reject candidates which do not satisfy this constraint.

Next, if \src{} contains loops, establishing the functional equivalence of the generated PS for all program states with \src{} requires loop invariants. In VL, loop invariants typically follow a templated structure: 
\begin{equation}
    Inv \; \triangleq \; f(i) \; \land \; e(T_{lang})
\end{equation} 
where $f(i)$ denotes an expression over loop indexes and $e(T_{lang})$ represents an expression recursively constructed using operators from \tlang{}. This structured nature simplifies the invariant generation process compared to solving general loop invariant synthesis problems. To facilitate the generation of loop invariants, we use 1-shot learning to familiarize the model with the concept and structure of invariants in the VL context (due to space constraints we illustrate in~\cref{sec:prompt}). The prompt for invariant generation closely resembles that used for generating program summaries, including \src{} with an additional assertion stating the equality of the return variable with the previously generated PS. This instruction guides the model to produce an invariant corresponding to the generated PS. 
The invariants are generated as Boolean expressions in Python rather than SMT-LIB, as we found that LLMs encounter difficulties in generating SMT-LIB (standard format for SMT-based theorem provers) code due to its limited representation in training datasets. When prompted, model generates the following invariant for the code shown in~\cref{fig:src}:
\begin{lstlisting}[language=python,numbers=none]
def invariant(data, i):
  return i >= 0 and i <= len(data) and
         sum = reduce(map(data[:i], lambda i : ite(i < 100, i, 0)),
                       lambda a, b: a + b)
\end{lstlisting}
The loop invariant states that the loop index $i$ remains within the bounds of the data array ($0 \leq i \leq len(data)$). Additionally, the invariant expresses $sum$ as the MapReduce expression over the first $i$ elements of the data array, which helps prove that the invariant holds in each iteration of the loop.

Both the program summaries and invariants are expressed in Python. We use simple pattern-matching rewrite rules to translate the expressions to the syntax compatible with the verification oracle used to check for functional equivalence. Once verified, the PS is similarly translated to the concrete syntax of \tlang{} using straightforward rewrite rules, leveraging the syntactic nature of Python. The translation process is simplified due to Python's highly structured syntax. We present our complete algorithm for generating PS and $Inv$ in ~\cref{sec:algo}.

\section{Experiments}
\label{sec:experiments}
To evaluate the effectiveness of \compiler, we evaluate across four distinct DSLs\footnote{All the benchmarks used for evaluation can be found at: https://drive.google.com/drive/folders/1vyxlREe8-gJ1BJviDN5tqMectwYMmcOr?usp=sharing}, each targeting diverse application domains: 

\begin{enumerate}[nosep,leftmargin=1.5em,labelwidth=*,align=left]
\item \textbf{Distributed Computing}: We transpile sequential Java programs into MapReduce implementations written using the Apache Spark~\cite{spark} API. Spark, an open-source distributed computing framework, provides an interface for programming multiple clusters which for data parallelism which helps in large-scale data processing.

\item \textbf{Network Packet Processing}: We transpile sequential network processing algorithms in C to the operators of programmable switch devices~\cite{domino} with its own ISA. This translation enables the exploration of novel algorithms, such as congestion control and load balancing, on programmable switch devices.

\item \textbf{TACO}: We transpile sequential C++ programs into TACO~\cite{taco}'s API. Taco is a tensor processing compiler for generating highly optimized GPU code for performing tensor computations.

\item \textbf{Tensor Processing}. We transpile sequential C++ programs to a tensor processing IR recently introduced by ~\cite{qiu2024tenspiler}. The tensor processing IR consists of common tensor operations such as element wise arithmetic operators, reduction operators and traspose, among others. This IR facilitates translation of unoptimized sequential code to tensor operations which can be then executed on 6 different software and hardware backends.
\end{enumerate}

\textbf{Model}: In all experiments, we use GPT-4 via their APIs to generate candidates. We set the temperature to 0.7 for all the experiments. For program summary and invariant generation across all domains, we use the same zero-shot PS prompt in \cref{fig:prompt_ps} and one-shot prompt in \cref{fig:prompt_inv}, respectively. We keep a budget of 50 queries for the PS and a budget of 10 queries for each PS. 

We present the results in the sections below and defer the error analysis to~\cref{app:error-analysis}.

\subsection{Distributed Computing}
MapReduce, a programming model for parallel processing of large datasets across distributed clusters, simplifies parallel computation by abstracting away distributed system complexities. It comprises two phases:
\begin {enumerate*}
\item Map: Input data is partitioned into smaller chunks, each processed by a mapper function to generate key-value pairs.
\item Reduce: Intermediate key-value pairs are shuffled, sorted based on keys, and then processed by reducer functions to aggregate associated values.
\end{enumerate*}

\textbf{\compiler implementation}. We compare the performance of \compiler against MetaLift~\cite{metalift}\footnote{Casper~\cite{casper} is not functional and 
Mold~\cite{mold} is not open-sourced}. MetaLift uses a symbolic solver (Rosette~\cite{rosette}) to perform the search. We evaluate on the same 45 benchmarks as MetaLift. All the benchmarks have loops and require loop invariants to prove the functional equivalence of the source and the generated program. MetaLift solves 40 out of 45 with a timeout of 1 hour. \compiler is able to solve \textbf{44}, i.e., generate the correct translation as well as the required  invariants to prove the correctness. \compiler solves 4 additional benchmarks on which MetaLift times out. In addition to solving more benchmarks, \compiler solves them much faster. It takes less than \textbf{1} minute on average to solve each benchmark when MetaLift has to take an average of 3 minutes to solve. The amount of effort required to build \compiler is alo significantly less than MetaLift as it does not require the developers to provide any search-space description for PS and invariants. Metalift requires over $\approx$ 1000 LoC for the description of these search-space.

\subsection{Network Packet Processing}
Network packet processing hardware, such as routers and switches, lacks flexibility post-development, preventing experimentation with new data-plane algorithms. Recently, a verified lifting approach ~\cite{domino} was introduced to simplify this process. This compiler offers the developers with two constructs:
\begin{enumerate*}
\item a packet transaction language (subset of the C language) to express the semantics of these data-plane algorithms
\item a compiler~\cite{domino} that translates the packet processing algorithms to the instruction set of programmable switch devices. 
\end{enumerate*}
Atoms are introduced as an instruction set of the hardware to represent the atomic operations supported by the hardware. Compiler translates the packet transaction algorithm to a sequence of atoms resulting in a different programmable switch configuration. 

\textbf{\compiler implementation}. We implement the Domino compiler using \compiler by defining the semantics of the atoms in the prompt. We compare the performance of our implementation against MetaLift’s implementation. All benchmarks in Domino are imperative C programs without any loop constructs, so no loop invariants are required for these benchmarks. The generated PS are verified using a SMT solver. MetaLift solves all the 10 benchmarks with an average time of 6 seconds. \compiler is also able to transpile all the \textbf{10} benchmarks but with an average time of only \textbf{2} seconds. Similar to the Spark case study, we do not require developers to specify the search-space for PS while Metalift requires over $\approx$ 1100 LoC to describe this search-space. In summary, \compiler shows similar performance to the existing compiler but can be built using much less effort. 

\subsection{TACO}
Tensors form the key construct in machine learning and tensor compilers play an important role in optimizing these operations. TACO~\cite{taco} is one such compiler which can automatically generate highly optimized code tailored to CPUs and GPUs. TACO’s language represents the operations in a concise Einsum like notation. Recently, C2TACO~\cite{c2taco} a search-based lifting tool was proposed to automate the translation of C++ code to TACO.

\textbf{\compiler implementation}. In~\cref{tab:taco_exp}, we compare the performance of C2TACO and \compiler for all the benchmarks. We use the same 90 mins timeout for each benchmark that was used in the original C2TACO evaluation~\cite{c2taco}. C2TACO solves 57 out of the total 60 benchmarks, while \compiler successfully solves all \textbf{60} benchmarks. The 3 benchmarks that C2TACO fails to solve require expressions of depth greater than 4. Due to its enumerative approach, C2TACO struggles to find solutions for these cases. We attempted to run these 3 challenging benchmarks with an extended timeout of 1 day, but the C2TACO solver was still unable to find a solution. C2TACO uses over 1000 LoC for implementing the heuristics to scale the symbolic search. In contrast, \compiler relies on a simple \textbf{100} lines of prompt (task instruction + DSL semantics) to achieve better performance than C2TACO. C2TACO takes an average of 41 seconds while \compiler average solving time is \textbf{2} seconds. We also perform an experiment to test the scalability of C2TACO enumerate apporach with more complex benchmarks than the ones used in the original evaluation. We include the results in ~\cref{sec:scale}.

\subsection{Tensor Processing}
Many domains, such as image processing, signal processing, and deep learning, have legacy code written in high-level languages that operate on individual values of the input and perform specific operations. To leverage the optimizations provided by deep learning frameworks or hardware backends like GPUs, this code needs to be lifted to the operators supported by these languages. Prior work by~\cite{qiu2024tenspiler} introduced a common tensor IR  that can translate sequential programs to six different hardware and software backends automatically using a verified lifting approach. 

\textbf{\compiler implementation}. We evaluate \compiler against Tenspiler~\cite{qiu2024tenspiler} on the 23 benchmarks from the image processing and ML kernel domain\footnote{We refer the readers to the paper~\cite{qiu2024tenspiler} for more details on these benchmarks.}. Tenspiler is able to solve all 23 benchmarks. \compiler also successfully solves all \textbf{23} benchmarks (including generating the correct proofs). However, it is important to note that Tenspiler's synthesis algorithm relies on three domain-specific optimizations to achieve scalability. These optimizations require significant effort to implement, with over $\approx$ 1200 LoC written by a domain expert. In contrast, \compiler solves these benchmarks without relying on any user-defined heuristics, showcasing its ability to generate correct solutions without the need for domain-specific optimizations. To check the scalability of Tenspiler's symblioc approach, we remove all the optimizations. Tenspiler without the optimizations can only solve 5 out of the 23 benchmarks with a timeout of 1 hour, highlighting the importance of the domain-specific optimizations for its performance. These results highlight the ability of \compiler to solve complex benchmarks without relying on domain-specific heuristics. Moreover, \compiler solves these benchmarks faster than Tenspiler with all its optimizations enabled. \compiler takes an average time of \textbf{95.89} seconds to solve each benchmark, whereas Tenspiler takes 115.14 seconds.

\subsection{Two-phase Approach for \compiler}
In this section, we evaluate an alternative approach to the two-phase method described in \cref{sec:approach}, where we generate the $Inv$(s) and the PS together in a single step. To test this, we prompt the model in a one-shot setting, providing an example that demonstrates generating the PS and the $Inv$(s) simultaneously. We merge the prompts described in \cref{fig:prompt_ps} and \cref{fig:prompt_inv} to create a unified prompt for this experiment.

Due to budget constraints, we limit this experiment to the tensor processing domain, which represents our most complex DSL with 37 operators. We use the same query budget as the two-phase approach. When prompted to generate the invariant and PS together, \compiler successfully solves 20 out of the total 23 benchmarks. In contrast, the two-phase approach described in \cref{sec:approach} solves all \textbf{23} benchmarks. We hypothesize that the reduced performance of the single-phase approach may be attributed to the increased complexity of generating both the PS and the $Inv$(s) simultaneously. Moreover, the two-phase approach enables the model to leverage the generated PS when constructing the invariant. By having access to the PS, the model can more effectively reason about the necessary conditions and constraints required for the invariant to hold.

\begin{table}
\footnotesize
    \centering
    \begin{tabular}{||ccccccccc||}
    \hline
        Tool   & BLAS  & DSP & DSPStone  & makespeare & mathfu  & simpl\_array & UTDSP & darknet \\
        \hline
        C2TACO &  100\% & 100\% & 100\% & 100\%  & 91.6\% & 90\% & 100\% & 92.8\%\\
        \hline
        \compiler  & 100\%  & 100\%  & 100\% & 100\%  & \textbf{100\%} & 100\%  & \textbf{100}\% & \textbf{100\%} \\
        \hline
    \end{tabular}
    \caption{Accuracy on various benchmarks for tensor processing domain.}
    \label{tab:taco_exp}
\end{table}

\section{Related Work}
\label{sec:related}
\textbf{Code Transpilation.} Several approaches have been proposed for automating the task of translating legacy or unoptimized code to DSLs. These range from symbolic rule-based approaches~\cite{mold} to search-based verified lifting approaches~\citep{casper, dexter, metalift, c2taco} and neural approaches~\cite{ngst,roziere2020unsupervised}. Most of these tools are either optimized for a specific domain or require domain expertise to scale. 
In contrast, \compiler simplifies the process of building lifting tools by leveraging LLMs.
The closest work to ours is~\cite{roziere2020unsupervised} which uses a sequence-to-sequence model to translate code between C++, Java, and Python; our work differs in two key respects: we target lifting to DSLs, and our LLM based approach produces formally verified code. 

\textbf{LLMs for Code.} LLMs are trained on massive amounts of code from various sources, leading to impressive performance on programming tasks such as code generation~\citep{starcoder,alphacode}, repair, testing, and transpilation. Furthermore, LLMs have been successfully employed to aid in certain formal methods tasks, including generating proofs and specifications~\citep{lemur,llminv,llminvrank}. However, generating reliable code from LLMs remains challenging due to the stochastic nature of these models and the lack of an external verification oracle. With \compiler, we demonstrate a novel approach to generating and verifying the generated code using LLMs.

\section{Conclusion}
We presented a principled approach to leverage LLMs for code transpilation. Unlike prior LLM-based transpilers, our transpiled code is {\em provably equivalent} to the input, while also takes significantly less time to generate as compared to prior non LLM-based approaches with correctness guarantee, as demonstrated in transpiling to 4 real-world DSLs.

\bibliography{main}

\begin{thebibliography}{26}
\providecommand{\natexlab}[1]{#1}
\providecommand{\url}[1]{\texttt{#1}}
\expandafter\ifx\csname urlstyle\endcsname\relax
  \providecommand{\doi}[1]{doi: #1}\else
  \providecommand{\doi}{doi: \begingroup \urlstyle{rm}\Url}\fi

\bibitem[Radoi et~al.(2014)Radoi, Fink, Rabbah, and Sridharan]{mold}
Cosmin Radoi, Stephen~J. Fink, Rodric Rabbah, and Manu Sridharan.
\newblock Translating imperative code to mapreduce.
\newblock In \emph{Proceedings of the 2014 ACM International Conference on Object Oriented Programming Systems Languages \& Applications}, OOPSLA '14, pages 909--927, New York, NY, USA, 2014. ACM.
\newblock ISBN 978-1-4503-2585-1.
\newblock \doi{10.1145/2660193.2660228}.

\bibitem[Solar-Lezama et~al.(2006)Solar-Lezama, Tancau, Bod\'{\i}k, Seshia, and Saraswat]{solar-asplos06}
Armando Solar-Lezama, Liviu Tancau, Rastislav Bod\'{\i}k, Sanjit~A. Seshia, and Vijay~A. Saraswat.
\newblock Combinatorial sketching for finite programs.
\newblock In \emph{Proceedings of the 12th International Conference on Architectural Support for Programming Languages and Operating Systems (ASPLOS)}, pages 404--415. ACM Press, October 2006.

\bibitem[Jha et~al.(2010)Jha, Gulwani, Seshia, and Tiwari]{jha-icse10}
Susmit Jha, Sumit Gulwani, Sanjit~A. Seshia, and Ashish Tiwari.
\newblock Oracle-guided component-based program synthesis.
\newblock In \emph{Proceedings of the 32nd ACM/IEEE International Conference on Software Engineering (ICSE)}, pages 215--224, May 2010.

\bibitem[Gulwani et~al.(2017)Gulwani, Polozov, and Singh]{progsynthesis-bookch17}
Sumit Gulwani, Oleksandr Polozov, and Rishabh Singh.
\newblock Program synthesis.
\newblock \emph{Found. Trends Program. Lang.}, 4\penalty0 (1-2):\penalty0 1--119, 2017.

\bibitem[Ahmad and Cheung(2018)]{casper}
Maaz Bin~Safeer Ahmad and Alvin Cheung.
\newblock Automatically leveraging mapreduce frameworks for data-intensive applications.
\newblock In Gautam Das, Christopher~M. Jermaine, and Philip~A. Bernstein, editors, \emph{Proceedings of the 2018 International Conference on Management of Data, {SIGMOD} Conference 2018, Houston, TX, USA, June 10-15, 2018}, pages 1205--1220. {ACM}, 2018.

\bibitem[Magalh\~{a}es et~al.(2023)Magalh\~{a}es, Woodruff, Polgreen, and O'Boyle]{c2taco}
Jos\'{e} Wesley de~Souza Magalh\~{a}es, Jackson Woodruff, Elizabeth Polgreen, and Michael F.~P. O'Boyle.
\newblock C2taco: Lifting tensor code to taco.
\newblock In \emph{Proceedings of the 22nd ACM SIGPLAN International Conference on Generative Programming: Concepts and Experiences}, GPCE 2023, page 42–56, New York, NY, USA, 2023. Association for Computing Machinery.
\newblock ISBN 9798400704062.
\newblock \doi{10.1145/3624007.3624053}.
\newblock URL \url{https://doi.org/10.1145/3624007.3624053}.

\bibitem[Cheung et~al.(2013)Cheung, Solar-Lezama, and Madden]{qbs}
Alvin Cheung, Armando Solar-Lezama, and Samuel Madden.
\newblock Optimizing database-backed applications with query synthesis.
\newblock \emph{ACM SIGPLAN Notices}, 48\penalty0 (6):\penalty0 3--14, 2013.

\bibitem[Ahmad et~al.(2019)Ahmad, Ragan-Kelley, Cheung, and Kamil]{dexter}
Maaz Bin~Safeer Ahmad, Jonathan Ragan-Kelley, Alvin Cheung, and Shoaib Kamil.
\newblock Automatically translating image processing libraries to halide.
\newblock \emph{ACM Transactions on Graphics (TOG)}, 38\penalty0 (6):\penalty0 1--13, 2019.

\bibitem[Mariano et~al.(2022)Mariano, Chen, Feng, Durrett, and Dillig]{ngst}
Benjamin Mariano, Yanju Chen, Yu~Feng, Greg Durrett, and I\c{s}il Dillig.
\newblock Automated transpilation of imperative to functional code using neural-guided program synthesis.
\newblock \emph{Proc. ACM Program. Lang.}, 6\penalty0 (OOPSLA1), April 2022.
\newblock \doi{10.1145/3527315}.
\newblock URL \url{https://doi.org/10.1145/3527315}.

\bibitem[Devlin et~al.(2019)Devlin, Chang, Lee, and Toutanova]{bert}
Jacob Devlin, Ming-Wei Chang, Kenton Lee, and Kristina Toutanova.
\newblock Bert: Pre-training of deep bidirectional transformers for language understanding.
\newblock In \emph{North American Chapter of the Association for Computational Linguistics}, 2019.
\newblock URL \url{https://api.semanticscholar.org/CorpusID:52967399}.

\bibitem[Brown et~al.(2020)Brown, Mann, Ryder, Subbiah, Kaplan, Dhariwal, Neelakantan, Shyam, Sastry, Askell, Agarwal, Herbert-Voss, Krueger, Henighan, Child, Ramesh, Ziegler, Wu, Winter, Hesse, Chen, Sigler, Litwin, Gray, Chess, Clark, Berner, McCandlish, Radford, Sutskever, and Amodei]{few-shot}
Tom~B. Brown, Benjamin Mann, Nick Ryder, Melanie Subbiah, Jared Kaplan, Prafulla Dhariwal, Arvind Neelakantan, Pranav Shyam, Girish Sastry, Amanda Askell, Sandhini Agarwal, Ariel Herbert-Voss, Gretchen Krueger, Tom Henighan, Rewon Child, Aditya Ramesh, Daniel~M. Ziegler, Jeffrey Wu, Clemens Winter, Christopher Hesse, Mark Chen, Eric Sigler, Mateusz Litwin, Scott Gray, Benjamin Chess, Jack Clark, Christopher Berner, Sam McCandlish, Alec Radford, Ilya Sutskever, and Dario Amodei.
\newblock Language models are few-shot learners.
\newblock In \emph{Proceedings of the 34th International Conference on Neural Information Processing Systems}, NIPS'20, Red Hook, NY, USA, 2020. Curran Associates Inc.
\newblock ISBN 9781713829546.

\bibitem[Li et~al.(2023)Li, Allal, Zi, Muennighoff, Kocetkov, Mou, Marone, Akiki, Li, Chim, Liu, Zheltonozhskii, Zhuo, Wang, Dehaene, Davaadorj, Lamy-Poirier, Monteiro, Shliazhko, Gontier, Meade, Zebaze, Yee, Umapathi, Zhu, Lipkin, Oblokulov, Wang, Murthy, Stillerman, Patel, Abulkhanov, Zocca, Dey, Zhang, Fahmy, Bhattacharyya, Yu, Singh, Luccioni, Villegas, Kunakov, Zhdanov, Romero, Lee, Timor, Ding, Schlesinger, Schoelkopf, Ebert, Dao, Mishra, Gu, Robinson, Anderson, Dolan-Gavitt, Contractor, Reddy, Fried, Bahdanau, Jernite, Ferrandis, Hughes, Wolf, Guha, von Werra, and de~Vries]{starcoder}
Raymond Li, Loubna~Ben Allal, Yangtian Zi, Niklas Muennighoff, Denis Kocetkov, Chenghao Mou, Marc Marone, Christopher Akiki, Jia Li, Jenny Chim, Qian Liu, Evgenii Zheltonozhskii, Terry~Yue Zhuo, Thomas Wang, Olivier Dehaene, Mishig Davaadorj, Joel Lamy-Poirier, João Monteiro, Oleh Shliazhko, Nicolas Gontier, Nicholas Meade, Armel Zebaze, Ming-Ho Yee, Logesh~Kumar Umapathi, Jian Zhu, Benjamin Lipkin, Muhtasham Oblokulov, Zhiruo Wang, Rudra Murthy, Jason Stillerman, Siva~Sankalp Patel, Dmitry Abulkhanov, Marco Zocca, Manan Dey, Zhihan Zhang, Nour Fahmy, Urvashi Bhattacharyya, Wenhao Yu, Swayam Singh, Sasha Luccioni, Paulo Villegas, Maxim Kunakov, Fedor Zhdanov, Manuel Romero, Tony Lee, Nadav Timor, Jennifer Ding, Claire Schlesinger, Hailey Schoelkopf, Jan Ebert, Tri Dao, Mayank Mishra, Alex Gu, Jennifer Robinson, Carolyn~Jane Anderson, Brendan Dolan-Gavitt, Danish Contractor, Siva Reddy, Daniel Fried, Dzmitry Bahdanau, Yacine Jernite, Carlos~Muñoz Ferrandis, Sean Hughes, Thomas Wolf, Arjun Guha, Leandro von
  Werra, and Harm de~Vries.
\newblock Starcoder: may the source be with you!, 2023.

\bibitem[Li et~al.(2022)Li, Choi, Chung, Kushman, Schrittwieser, Leblond, Eccles, Keeling, Gimeno, Dal~Lago, Hubert, Choy, de~Masson~d’Autume, Babuschkin, Chen, Huang, Welbl, Gowal, Cherepanov, Molloy, Mankowitz, Sutherland~Robson, Kohli, de~Freitas, Kavukcuoglu, and Vinyals]{alphacode}
Yujia Li, David Choi, Junyoung Chung, Nate Kushman, Julian Schrittwieser, Rémi Leblond, Tom Eccles, James Keeling, Felix Gimeno, Agustin Dal~Lago, Thomas Hubert, Peter Choy, Cyprien de~Masson~d’Autume, Igor Babuschkin, Xinyun Chen, Po-Sen Huang, Johannes Welbl, Sven Gowal, Alexey Cherepanov, James Molloy, Daniel~J. Mankowitz, Esme Sutherland~Robson, Pushmeet Kohli, Nando de~Freitas, Koray Kavukcuoglu, and Oriol Vinyals.
\newblock Competition-level code generation with alphacode.
\newblock \emph{Science}, 378\penalty0 (6624):\penalty0 1092–1097, December 2022.
\newblock ISSN 1095-9203.
\newblock \doi{10.1126/science.abq1158}.
\newblock URL \url{http://dx.doi.org/10.1126/science.abq1158}.

\bibitem[Rozière et~al.(2024)Rozière, Gehring, Gloeckle, Sootla, Gat, Tan, Adi, Liu, Sauvestre, Remez, Rapin, Kozhevnikov, Evtimov, Bitton, Bhatt, Ferrer, Grattafiori, Xiong, Défossez, Copet, Azhar, Touvron, Martin, Usunier, Scialom, and Synnaeve]{codellama}
Baptiste Rozière, Jonas Gehring, Fabian Gloeckle, Sten Sootla, Itai Gat, Xiaoqing~Ellen Tan, Yossi Adi, Jingyu Liu, Romain Sauvestre, Tal Remez, Jérémy Rapin, Artyom Kozhevnikov, Ivan Evtimov, Joanna Bitton, Manish Bhatt, Cristian~Canton Ferrer, Aaron Grattafiori, Wenhan Xiong, Alexandre Défossez, Jade Copet, Faisal Azhar, Hugo Touvron, Louis Martin, Nicolas Usunier, Thomas Scialom, and Gabriel Synnaeve.
\newblock Code llama: Open foundation models for code, 2024.

\bibitem[Pei et~al.(2023)Pei, Bieber, Shi, Sutton, and Yin]{llminv}
Kexin Pei, David Bieber, Kensen Shi, Charles Sutton, and Pengcheng Yin.
\newblock Can large language models reason about program invariants?
\newblock In Andreas Krause, Emma Brunskill, Kyunghyun Cho, Barbara Engelhardt, Sivan Sabato, and Jonathan Scarlett, editors, \emph{Proceedings of the 40th International Conference on Machine Learning}, volume 202 of \emph{Proceedings of Machine Learning Research}, pages 27496--27520. PMLR, 23--29 Jul 2023.
\newblock URL \url{https://proceedings.mlr.press/v202/pei23a.html}.

\bibitem[Chakraborty et~al.(2023)Chakraborty, Lahiri, Fakhoury, Musuvathi, Lal, Rastogi, Senthilnathan, Sharma, and Swamy]{llminvrank}
Saikat Chakraborty, Shuvendu~K Lahiri, Sarah Fakhoury, Madanlal Musuvathi, Akash Lal, Aseem Rastogi, Aditya Senthilnathan, Rahul Sharma, and Nikhil Swamy.
\newblock Ranking llm-generated loop invariants for program verification.
\newblock \emph{arXiv preprint arXiv:2310.09342}, 2023.

\bibitem[Alur et~al.(2013)Alur, Bodik, Juniwal, Martin, Raghothaman, Seshia, Singh, Solar-Lezama, Torlak, and Udupa]{sygus}
Rajeev Alur, Rastislav Bodik, Garvit Juniwal, Milo M.~K. Martin, Mukund Raghothaman, Sanjit~A. Seshia, Rishabh Singh, Armando Solar-Lezama, Emina Torlak, and Abhishek Udupa.
\newblock Syntax-guided synthesis.
\newblock In \emph{2013 Formal Methods in Computer-Aided Design}, pages 1--8, 2013.
\newblock \doi{10.1109/FMCAD.2013.6679385}.

\bibitem[Kjolstad et~al.(2017)Kjolstad, Kamil, Chou, Lugato, and Amarasinghe]{taco}
Fredrik Kjolstad, Shoaib Kamil, Stephen Chou, David Lugato, and Saman Amarasinghe.
\newblock The tensor algebra compiler.
\newblock \emph{Proc. ACM Program. Lang.}, 1\penalty0 (OOPSLA):\penalty0 77:1--77:29, October 2017.
\newblock ISSN 2475-1421.
\newblock \doi{10.1145/3133901}.
\newblock URL \url{http://doi.acm.org/10.1145/3133901}.

\bibitem[Hoare(1969)]{hoare-logic}
C.~A.~R. Hoare.
\newblock An axiomatic basis for computer programming.
\newblock \emph{Commun. {ACM}}, 12\penalty0 (10):\penalty0 576--580, 1969.

\bibitem[Zaharia et~al.(2012)Zaharia, Chowdhury, Das, Dave, Ma, McCauley, Franklin, Shenker, and Stoica]{spark}
Matei Zaharia, Mosharaf Chowdhury, Tathagata Das, Ankur Dave, Justin Ma, Murphy McCauley, Michael~J. Franklin, Scott Shenker, and Ion Stoica.
\newblock Resilient distributed datasets: A fault-tolerant abstraction for in-memory cluster computing.
\newblock In \emph{Proceedings of the 9th USENIX Conference on Networked Systems Design and Implementation}, NSDI'12, 2012.

\bibitem[Sivaraman et~al.(2016)Sivaraman, Cheung, Budiu, Kim, Alizadeh, Balakrishnan, Varghese, McKeown, and Licking]{domino}
Anirudh Sivaraman, Alvin Cheung, Mihai Budiu, Changhoon Kim, Mohammad Alizadeh, Hari Balakrishnan, George Varghese, Nick McKeown, and Steve Licking.
\newblock Packet transactions: High-level programming for line-rate switches.
\newblock In \emph{Proceedings of the {ACM} {SIGCOMM} 2016 Conference, Florianopolis, Brazil, August 22-26, 2016}, pages 15--28, 2016.

\bibitem[Qiu et~al.(2024)Qiu, Cai, Bhatia, Hasabnis, Seshia, and Cheung]{qiu2024tenspiler}
Jie Qiu, Colin Cai, Sahil Bhatia, Niranjan Hasabnis, Sanjit~A. Seshia, and Alvin Cheung.
\newblock Tenspiler: A verified lifting-based compiler for tensor operations, 2024.

\bibitem[Bhatia et~al.(2023)Bhatia, Kohli, Seshia, and Cheung]{metalift}
Sahil Bhatia, Sumer Kohli, Sanjit~A. Seshia, and Alvin Cheung.
\newblock {Building Code Transpilers for Domain-Specific Languages Using Program Synthesis}.
\newblock In Karim Ali and Guido Salvaneschi, editors, \emph{37th European Conference on Object-Oriented Programming (ECOOP 2023)}, volume 263 of \emph{Leibniz International Proceedings in Informatics (LIPIcs)}, pages 38:1--38:30, Dagstuhl, Germany, 2023. Schloss Dagstuhl -- Leibniz-Zentrum f{\"u}r Informatik.
\newblock ISBN 978-3-95977-281-5.
\newblock \doi{10.4230/LIPIcs.ECOOP.2023.38}.
\newblock URL \url{https://drops.dagstuhl.de/entities/document/10.4230/LIPIcs.ECOOP.2023.38}.

\bibitem[Torlak and Bodik(2013)]{rosette}
Emina Torlak and Rastislav Bodik.
\newblock Growing solver-aided languages with rosette.
\newblock In \emph{Proceedings of the 2013 ACM International Symposium on New Ideas, New Paradigms, and Reflections on Programming \& Software}, Onward! 2013, pages 135--152, New York, NY, USA, 2013. ACM.
\newblock ISBN 978-1-4503-2472-4.
\newblock \doi{10.1145/2509578.2509586}.
\newblock URL \url{http://doi.acm.org/10.1145/2509578.2509586}.

\bibitem[Roziere et~al.(2020)Roziere, Lachaux, Chanussot, and Lample]{roziere2020unsupervised}
Baptiste Roziere, Marie-Anne Lachaux, Lowik Chanussot, and Guillaume Lample.
\newblock Unsupervised translation of programming languages.
\newblock \emph{Advances in neural information processing systems}, 33:\penalty0 20601--20611, 2020.

\bibitem[Wu et~al.(2023)Wu, Barrett, and Narodytska]{lemur}
Haoze Wu, Clark Barrett, and Nina Narodytska.
\newblock Lemur: Integrating large language models in automated program verification.
\newblock \emph{arXiv preprint arXiv:2310.04870}, 2023.

\end{thebibliography}
\bibliographystyle{unsrtnat}
\newpage
\appendix
\noindent{\Large\bf Appendix}
\addcontentsline{toc}{section}{Appendices}

\section{Algorithm}
The {\tt transpile\_code} algorithm translates the source code of a program into a target language using LLM. The {\tt source\_code} parameter represents the code to be transpiled, and {\tt num\_iters} and {\tt n} specify the number of iterations and the number of PS and $Inv$s to generate in each iteration, respectively.

The algorithm maintains two sets: {\tt incorrect\_ps\_sols} and {\tt seen\_ps\_sols} (lines 2, 3). The {\tt incorrect\_ps\_sols} set keeps track of PS that are syntactically correct but have been found to be incorrect during the transpilation process. The {\tt seen\_ps\_sols} set keeps track of all PS that have been processed by the algorithm. The algorithm operates in a loop that runs for {\tt num\_iters} iterations (line 5). In each iteration, the algorithm calls {\tt get\_ps\_sols} to generate {\tt n} different PS for the given {\tt source\_code} from LLM, and supply to LLM any PS that have been marked as incorrect as stored in the {\tt incorrect\_ps\_sols} set (line 6). For each generated PS, the algorithm first checks if it has been seen before by looking it up in the {\tt seen\_ps\_sols} set (line 9). If the PS has been encountered previously, the algorithm skips it to avoid redundant processing. If it is new, the algorithm parses it to check for syntactic validity using the parse function (line 12). If the summary has invalid syntax, it is discarded, and the algorithm moves on to the next summary. If the PS is syntactically correct, the algorithm proceeds to generate $Inv$ for it. It maintains a set called {\tt seen\_inv\_sols\_for\_ps} that keeps track of invariants that have been processed for the current PS to avoid redundant processing. It also checks each generated $Inv$’s syntactic validity and discards it if it is not.

If both the PS and the $Inv$s pass the syntactic validation, the algorithm proceeds to verify their correctness using the {\tt verify} function (line 26). If the verification succeeds, the algorithm returns the PS ({\tt ps\_sol}) as the final transpiled code.
If none of the generated $Inv$s for a PS are found to be valid, the algorithm assumes that the PS itself is incorrect. It adds the PS to the {\tt incorrect\_ps\_sols} set to exclude it from future iterations and adds it to the {\tt seen\_ps\_sols} set to mark it as processed.
The algorithm continues this process of generating PS and $Inv$s and verifying their correctness until a valid solution is found or the maximum number of tries is reached. If no valid solution is found within the given number of tries, the algorithm returns None, indicating that the transpilation was unsuccessful.

\label{sec:algo}
\begin{figure}
\begin{lstlisting}[language=python,basicstyle=\scriptsize\ttfamily]
def transpile_code(source_code: str, num_iters: int, n: int) -> str:
    incorrect_ps_sols: set[str] = set()
    seen_ps_sols: set[str] = set()
    
    for _ in range(num_iters):
        ps_sols: list[str] = get_ps_sols(n, source_code, incorrect_ps_sols)
        for ps_sol in ps_sols:
            # We have processed this PS before
            if ps_sol in seen_ps_sols:
                continue
            # If this PS has invalid syntax
            if not parse(ps_sol):
                continue

            # Generate invariants for this PS
            seen_inv_sols_for_ps: set[str] = set()
            inv_sols: list[str] = get_inv_sols_for_ps(n, ps_sol)
            for inv_sol in inv_sols:
                # We have processed this INV for this PS before
                if inv_sol in seen_inv_sols_for_ps:
                    continue
                # If this INV has invalid syntax
                if not parse(inv_sol):
                    continue
                # Verify INV and PS.
                if verify(inv_sol, ps_sol):
                    return ps_sol
                seen_inv_sols_for_ps.add(inv_sol)

            # At this point, none of the INVs work for this PS. We assume this ps is incorrect.
            incorrect_ps_sols.add(ps_sol)
            seen_ps_sols.add(ps_sol)
            
    
    # No solution has been found.
    return None
\end{lstlisting}
\end{figure}

\section{Prompts}
\label{sec:prompt}
\begin{figure}
    \centering
    \includegraphics[scale=0.7, trim=0cm 0cm 0cm 0cm]{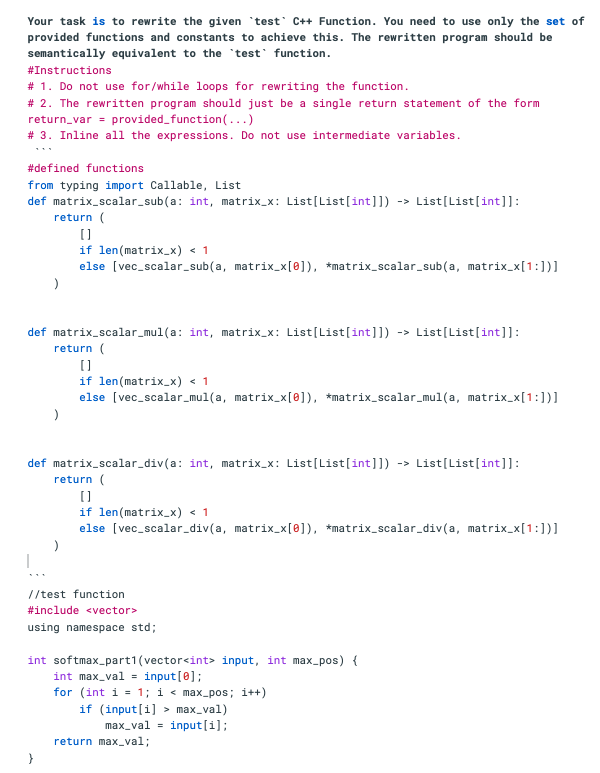}
    \caption{Program summary guessing prompt}
    \label{fig:prompt_ps}
\end{figure}

In this section, we present an instantiation of the prompt structure shown in \cref{fig:prompt_struc}. The prompt shown in ~\cref{fig:prompt_ps} consists of several components designed to guide the language model in generating semantically equivalent code using a restricted set of functions and constants. The prompt begins with a clear task description that instructs the model that its goal is to rewrite the given C++ function using only the provided functions and constants while maintaining semantic equivalence. Next, the prompt includes a set of instructions. These constraints are designed to make the generated code easier to parse and translate into a format suitable for theorem provers. The prompt then provides a set of defined functions in Python. These functions define all DSL operators that the model can use to rewrite the given C++ function. Finally, the prompt includes the test function in C++ which the model should rewrite using the provided functions and constants.

\begin{figure}[!ht]
\centering
\includegraphics[scale=0.20, trim=0cm 0cm 0cm 0cm]{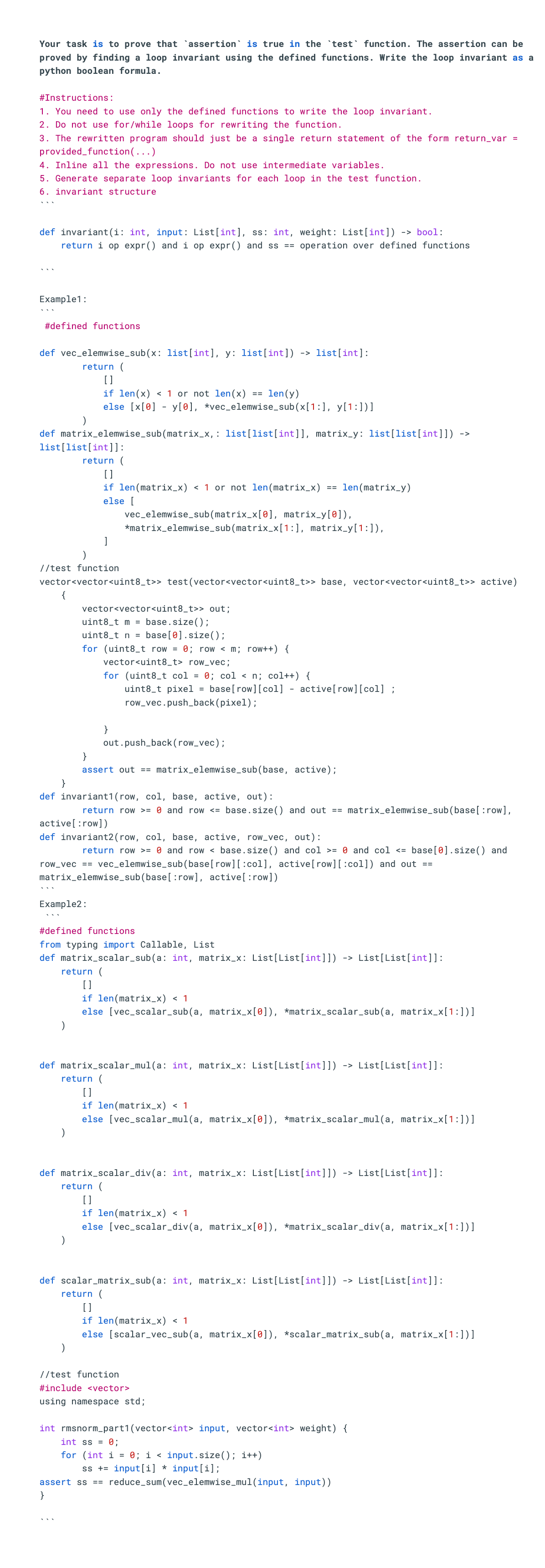}
\caption{Invariant guessing prompt}
\label{fig:prompt_inv}
\end{figure}

In ~\cref{fig:prompt_inv}, we present a one-shot prompt designed to guide the language model in generating loop invariants for the given {\tt test} function. This prompt is similar in structure to the program summary guessing prompt: it provides a clear task description, a set of instructions, and examples to guide the model in generating the desired output. The prompt instructs the model to prove the assertion in the {\tt test} function by finding a loop invariant using the defined functions. It includes specific constraints on the generated loop invariant, such as using only the defined functions, avoiding loops, using a single return statement, inlining expressions, and generating separate invariants for each loop in the {\tt test} function. These constraints are intended to simplify the parsing of the generated invariants into SMT formulas, making it easier to integrate them into automated theorem provers. Additionally, the prompt provides a template for the invariant structure, guiding the model in constructing the loop invariant as a Python function that takes the loop variables and relevant data structures as input and returns a boolean expression. The invariant should involve comparisons of the loop variables using operators and expressions, and an equality check for the loop-dependent variable using the defined functions. The prompt also includes an example to demonstrate the expected format and structure of the loop invariant. Example 1 shows a {\tt test} function that performs element-wise subtraction of two matrices and provides two loop invariants. Example 2 shows the {\tt test} function for which we need to generate the loop invariants.

To avoid regenerating the same incorrect solutions, the prompt also includes all the syntactically correct solutions that have been generated so far, along with a message saying \textit{These generated programs are incorrect. Do not generate the same. Please generate another program.}

\section{Scalability}
\label{sec:scale}
In  this section, we evaluate the scalability of symbolic solvers in the context of VL-based tools. The benchmarks used in the evaluation of these tools are often carefully selected and limited in scope, allowing the tools to perform well within their intended domain. However, in this experiment, we aim to demonstrate that symbolic tools relying on domain-specific heuristics can be brittle and fail to scale when the complexity of the benchmarks increases beyond a certain threshold.  

We begin by evaluating C2TACO. Upon careful analysis of the benchmarks on which C2TACO struggles, we observed that the tool often times out when tasked with generating expressions of length greater than 4 One such example is illustrated in \cref{fig:taco_fail} where the source performs an in-place operation on an array {\tt arr} of length n and raises each element of the array to the power of 4. C2TACO enumerates candidate expressions using tensor operators and index variables in increasing order of expression length.  C2TACO enumerates $\approx$\textbf{30k} candidates. We illustrate some of the incorrect expressions in ~\cref{fig:taco_fail}.

To test the scalability of C2TACO, we randomly generated a set of 10 benchmarks with expressions of varying lengths, ranging from 5 to 10, incorporating various arithmetic operations (see ~\cref{fig:taco_synth} for an example). We used a timeout of 90 minutes for C2TACO, as reported in the original evaluation. C2TACO was unable to solve any of the 10 benchmarks within the timeout. In contrast, \compiler,  was able to solve all \textbf{10} benchmarks correctly in less than \textbf{2 seconds}. This performance can be attributed to the ability of language models to identify patterns and learn from the context provided in the source code. To further test the capabilities of \compiler, we evaluated it on a variation of the benchmark shown in \cref{fig:taco_fail}, where each element of the array is raised to the power of 20 instead of 4. Despite the increased complexity of the expression, \compiler was able to generate the correct solution efficiently.

\begin{figure}
\begin{subfigure}{0.5\textwidth}
\begin{lstlisting}[language=cpp,basicstyle=\scriptsize\ttfamily]
void fourth_in_place(int* arr, int n)
{
  for (int i = 0; i < n; ++i) {
    arr[i] = arr[i] * arr[i];
    arr[i] = arr[i] * arr[i];
  }
}
//TACO expression
out[i] =  arr[i] * arr[i] * arr[i] * arr[i]

//Incorrect TACO expressions
out(i) = arr(i) * arr(i)
out(i) = Cons1 + arr(j,i)
out(k) = Cons1 * arr(l,k,j)
\end{lstlisting}
\caption{Benchmark on which C2TACO fails.}
\label{fig:taco_fail}
\end{subfigure}
\begin{subfigure}{0.5\textwidth}
\begin{lstlisting}[language=cpp,basicstyle=\scriptsize\ttfamily]
void test1(int* arr, int n)
{
  for (int i = 0; i < n; ++i) {
    arr[i] = arr[i] + arr[i] + arr[i] + arr[i] + arr[i];
  }
}
//TACO expression
out[i] =  arr[i] + arr[i] + arr[i] + arr[i] + arr[i]
\end{lstlisting}

\caption{Example of synthetic benchmark with expression length $=$ 5.}
\label{fig:taco_synth}
\end{subfigure}
\caption{Scalability evaluation of symbolic solvers using synthetic benchmarks}
\end{figure}

\section{Qualitative Analysis of the Errors.} 
\label{app:error-analysis}
In this section, we provide a qualitative analysis of the mistakes made by LLMs while generating code and proofs. In \compiler, we use Python as the IR and the ps and inv(s) are generated in Python. The errors encountered can be classified into two categories: syntactic and semantic.

Syntactic errors occur when the generated code constructs are not compatible with the theorem prover. To mitigate this issue, we use a syntactic parser that translates the generated solutions to the language supported by the theorem prover. The parser ensures that only supported constructs are present in the solutions and rejects any candidates that do not comply with the theorem prover's syntax.

One common source of syntactic errors is the use of Python-specific constructs that are not supported by SMT solvers. Although we prompt the model to generate solutions using only the constructs provided in the prompt's scope, controlling the exact code generated by the model can be challenging. \Cref{fig:syntactic_errors} illustrates examples of program summaries generated by GPT-4 for the {\tt screen blend} benchmark that contain unsupported constructs. For instance, the first solution in \cref{fig:syntactic_errors} uses a {\tt for} loop, which is not supported by SMT solvers. Similarly, the second and third solutions utilize Python's list comprehension syntax, which is also not directly supported by SMT solvers. List comprehension are supported in SMT solvers using empty lists and append functions, such as {\tt append(1, [])}.

Semantic errors occur when the generated code is syntactically correct but is semantically not equivalent to the given \src{}. In the context of the {\tt screen blend} benchmark (shown in \cref{fig:screen_src}), \cref{fig:semantic_errors} illustrates two examples of semantically incorrect programs generated by GPT-4.
The first program incorrectly subtracts a term from the {\tt base} matrix instead of subtracting it from the sum of {\tt base} and {\tt active} matrices.
The second program suffers from a similar issue. It subtracts an incorrect term from the {\tt active} matrix. Specifically, the term being subtracted is {\tt matrix\_elemwise\_div(matrix\_elemwise\_mul(base, active), matrix\_scalar\_mul(32, matrix\_elemwise\_mul(base, active)))}, which is different from the one in the given  program. 

\begin{figure}
\begin{lstlisting}[language=cpp,basicstyle=\scriptsize\ttfamily]
vector<vector<int>> screen_blend_8(vector<vector<int>> base, vector<vector<int>> active)
{
    vector<vector<int>> out;
    int m = base.size();
    int n = base[0].size();
	for (int row = 0; row < m; row++) {
        vector<int> row_vec;
		for (int col = 0; col < n; col++) {
			int pixel = base[row][col] + active[row][col] - (base[row][col] * active[row][col]) / 255;
			row_vec.push_back(pixel);
		}
		out.push_back(row_vec);
	}
	return out;
}
\end{lstlisting}
\caption{Screen\_blend benchmark source code.}
\label{fig:screen_src}
\end{figure}

\begin{figure}
\begin{subfigure}{\textwidth}
\begin{lstlisting}[language=python,basicstyle=\scriptsize\ttfamily]
def screen_blend_8(base: List[List[int]], active: List[List[int]]) -> List[List[int]]:
    return matrix_elemwise_add(
        matrix_elemwise_sub(
            base, 
            matrix_elemwise_div(
                matrix_elemwise_mul(base, active), 
                vec_scalar_mul(32, [1 for _ in range(len(base[0]))])
            )
        ),
        active
    )
\end{lstlisting}
\end{subfigure}
\begin{subfigure}{\textwidth}
\begin{lstlisting}[language=python,basicstyle=\scriptsize\ttfamily]
def screen_blend_8(base: List[List[int]], active: List[List[int]]) -> List[List[int]]:
    return matrix_elemwise_add(
        matrix_elemwise_sub(
            base,
            matrix_elemwise_div(
                matrix_elemwise_mul(base, active),
                vec_scalar_mul(32, [[1]*len(base[0])]*len(base))
            )
        ),
        active
    )
\end{lstlisting}
\end{subfigure}
\begin{subfigure}{\textwidth}
\begin{lstlisting}[language=python,basicstyle=\scriptsize\ttfamily]
def screen_blend_8(base: List[List[int]], active: List[List[int]]) -> List[List[int]]:
    return matrix_elemwise_add(
        matrix_elemwise_sub(
            base,
            matrix_elemwise_div(
                matrix_elemwise_mul(base, active),
                vec_scalar_mul(32, vec_elemwise_mul([1], [1]))
            )
        ),
        active
    )
\end{lstlisting}
\end{subfigure}
\caption{Programs rejected by \compiler's syntactic parser}
\label{fig:syntactic_errors}
\end{figure}

\begin{figure}
\begin{subfigure}{\textwidth}
\begin{lstlisting}[language=python,basicstyle=\scriptsize\ttfamily]
def screen_blend_8(base: List[List[int]], active: List[List[int]]) -> List[List[int]]:
    return matrix_elemwise_add(
        matrix_elemwise_sub(
            base,
            matrix_elemwise_div(
                matrix_elemwise_mul(base, active),
                scalar_matrix_div(32, base)
            )
        ),
        active
    )
\end{lstlisting}
\end{subfigure}
\begin{subfigure}{\textwidth}
\begin{lstlisting}[language=python,basicstyle=\scriptsize\ttfamily]
def screen_blend_8(base: List[List[int]], active: List[List[int]]) -> List[List[int]]:
    return matrix_elemwise_add(
        base, 
        matrix_elemwise_sub(
            active, 
            matrix_elemwise_div(
                matrix_elemwise_mul(base, active), 
                matrix_scalar_mul(32, matrix_elemwise_mul(base, active))
            )
        )
    )
\end{lstlisting}
\end{subfigure}
\caption{Programs rejected by theorem prover for semantic incorrectness}
\label{fig:semantic_errors}
\end{figure}

\end{document}